\documentclass[12pt]{iopart}
\usepackage{inputenc}
\usepackage{graphicx}
\usepackage{subfigure}
\usepackage{color}
\usepackage{cite}
\usepackage{setspace}
\usepackage{cases}
\usepackage{bm}

\renewcommand{\vec}[1]{\bm{#1}}

\begin{document}

\title[Repeated output coupling of ultracold Feshbach molecules from a Cs BEC]{Repeated output coupling of ultracold Feshbach molecules from a Cs BEC}

\author{M. P. K\"oppinger, P. D. Gregory, D. L. Jenkin, \\D. J. McCarron\footnote{Current address: Department of Physics, Yale University, New Haven, CT 06511, USA}, A. L. Marchant and S. L. Cornish}

\address{Joint Quantum Centre (JQC), Durham - Newcastle, Department of Physics, Durham University, Durham DH1 3LE, United Kingdom}
\ead{s.l.cornish@durham.ac.uk}

\begin{abstract}

We investigate magnetoassociation of ultracold Feshbach molecules from a Bose-Einstein condensate of Cs atoms and explore the spectrum of weakly bound molecular states close to the atomic threshold. By exploiting the variation of magnetic field experienced by a molecular cloud falling in the presence of a magnetic field gradient, we demonstrate the repeated output coupling of molecules from a single atomic cloud using a Feshbach resonance at 19.89\,G. Using this method we are able to produce up to 24 separate pulses of molecules from a single atomic condensate, with a molecular pulse created every 7.2~ms. Furthermore, by careful control of the magnetic bias field and gradient we are able to utilise an avoided crossing in the bound state spectrum at 13.3\,G to demonstrate exquisite control over the dynamics of the molecular clouds.

\end{abstract}

\maketitle


\section{Introduction}\label{sec:Introduction}

The creation and manipulation of ultracold molecules is the subject of much current research~\cite{Doyle2004, Hutson2006}. The rich internal energy level structure of the molecules, combined with precise control of the motional degrees of freedom, offers exciting prospects in the fields of ultracold chemistry~\cite{Ospelkaus2010}, precision measurement~\cite{Hudson2011}, quantum computing~\cite{Demille2002} and simulation of many-body systems~\cite{Micheli2006, Georgescu2014}. Direct cooling of molecules into the ultracold regime is challenging, though recent progress on laser cooling shows tremendous promise~\cite{Barry2014, Zhelyazkova2013, Hummon2013}. An alternative strategy is to exploit the efficient techniques for cooling atomic gases before associating ultracold atoms to create ultracold molecules. In this context, magnetoassociation of weakly bound molecules using Feshbach resonances has proved pivotal~\cite{Chin2010,Kohler2006}.
%

Feshbach resonances arise when the energy of two colliding atoms matches that of a quasi-bound molecular state. Coupling between the atomic and molecular states leads to an avoided crossing in the energy spectrum as a function of an applied magnetic field, and a resonance feature in the $s$-wave scattering length. Sweeping the magnetic field across the Feshbach resonance so as to adiabatically follow the avoided crossing can be used to populate the weakly bound molecular state responsible for the resonance. In reality, ultracold atomic gases are subject to considerable momentum spread which leads to a continuum of atomic scattering states and can limit the efficiency of the molecule production~\cite{Bogdan2003, Kohler2006}. Nevertheless, magnetoassociation was first used successfully to create ultracold Cs$_2$~\cite{Herbig2003} and $^{40}$K$_2$ molecules~\cite{Regal2003} in 2003, and has subsequently been investigated in a  range of different atomic systems~\cite{Chin2010}. The technique has been used to produce the first molecular condensate~\cite{Jochim2003a, Greiner2003a} and has proved an important first step in the production of molecules in the rovibrational ground state via stimulated Raman adiabatic passage (STIRAP)~\cite{Ni2008,Lang2008,Danzl2008}.

In this paper we investigate magnetoassociation in an ultracold atomic gas of Cs. In particular we demonstrate and characterise a technique which allows the repeated output coupling of ultracold molecules from a single trapped atomic cloud. Additionally we exploit the near threshold bound state structure of Cs$_2$ to demonstrate control of the dynamics of the molecular cloud, opening up a route to the accumulation of multiple molecular pulses in a  second trap and the realisation of a low-energy molecular collider.

\section{Cs scattering behaviour at low magnetic fields}\label{sec:ScatteringBehaviour}

\begin{figure}
	\centering
\includegraphics[width=0.8\textwidth]{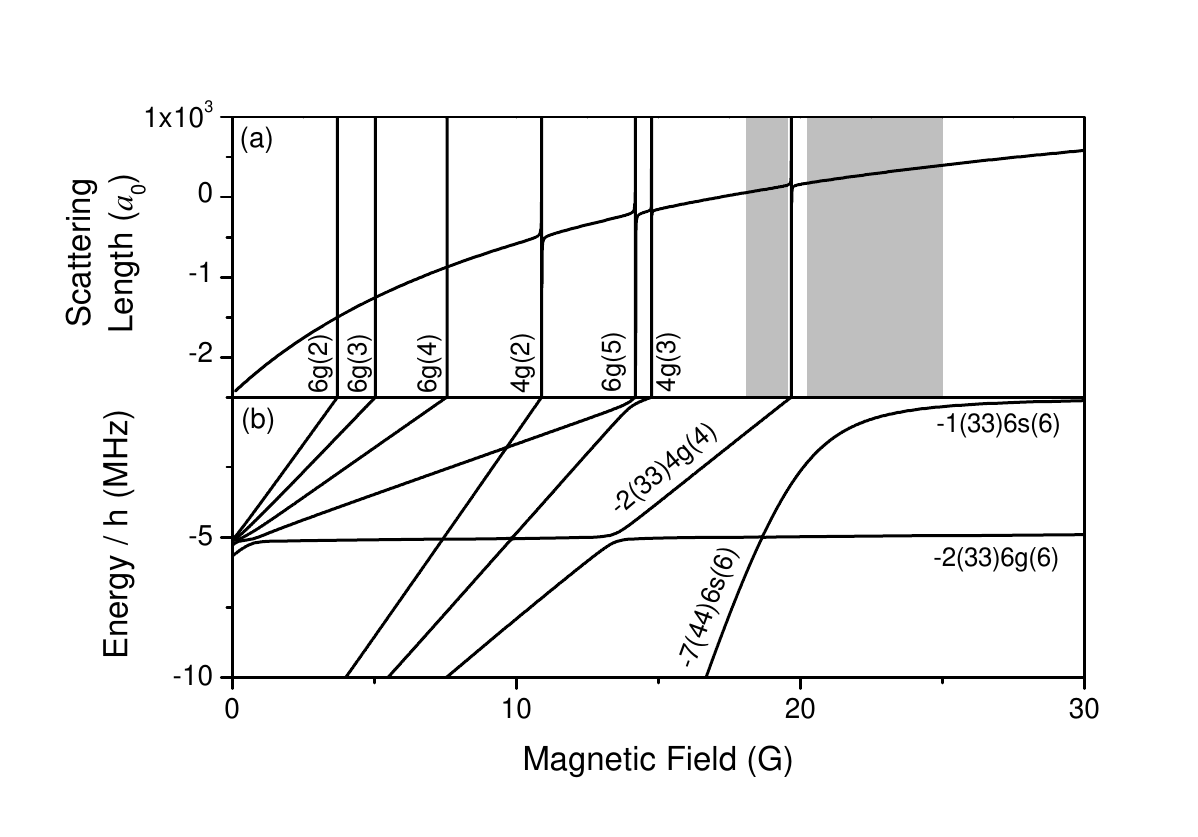}
	\caption[Cs scattering length and molecular bound states up to 30\,G.] {Cs scattering length and bound state up to 30\,G. (a) Scattering length of Cs in the $(f=3,m_f = +3)$ state. The grey areas mark the magnetic field regions where we are able to produce Cs Bose--Einstein condensates. (b) Cs bound state energies for weakly bound states. The notation is $n(f_1f_2)FL(M_F)$ where $n$ is the vibrational level counted from the least bound state, $f_i$ are the zero-field levels of the Cs atoms, $F$ is the resultant of the $f_i$, $L$ is the partial-wave angular momentum and $M_F=m_1+m_2$ following \cite{Berninger2013}.}
	\label{fig:CsScatBound30G}
\end{figure}

The experiments described in this paper use Bose--Einstein condensates of Cs in the $(f=3, m_f=+3)$ state confined in an optical trap. The use of this internal state eliminates the large two-body inelastic losses associated with the magnetically trappable states~\cite{Guery-Odelin1998,Thomas2003} and has proved essential for efficient cooling of Cs into the quantum degenerate regime~\cite{Weber2003}. Fortunately the magnetic field dependence of the scattering length in this state is highly favourable, both for tuning the elastic collision cross section and for magnetoassociation, as depicted in figure~\ref{fig:CsScatBound30G}. Here, the set of quantum numbers used for the near-threshold levels are $n(f_1f_2)FL(M_F)$, where $n$ is the vibrational level counted from the least bound state, $f_i$ are the zero-field levels of the Cs atoms, $F$ is the resultant of the $f_i$, $L$ is the partial-wave angular momentum and $M_F=m_1+m_2$ following \cite{Berninger2013}. In the rest of this paper, an abbreviation of the assignment to $FL(M_F)$ is used.

In the low field range, shown in figure~\ref{fig:CsScatBound30G}~(a), the scattering length varies markedly with magnetic field, changing from large and negative to large and positive, crossing zero at 17~G. The first Cs BEC~\cite{Weber2003} was realised close to this zero crossing where tuning of the scattering length to a moderate positive value was essential for efficient evaporation of the atomic sample. Subsequently, Cs$_2$ molecules~\cite{Mark2005} were produced using a Feshbach resonance at 19.89~G, allowing the study of the bound state spectrum~\cite{Mark2007a,Mark2007,Lange2009,Berninger2013} shown in figure 1(b). In these experiments, the Cs dimers were transferred into a different molecular state using an avoided crossing at 13.3~G. Transfer into this new state led to the observation of Feshbach-like resonance features \cite{Chin2005} between colliding Cs$_2$ molecules. In addition, Cs$_{2}$ Feshbach molecules have been transferred into a metastable state above the dissociation threshold \cite{Knoop2008}, and collisions between dimers \cite{Chin2005,Ferlaino2008,Ferlaino2010}, and between dimers and atoms \cite{Knoop2010} have been investigated. Most recently, Cs$_2$ Feshbach molecules have served as the starting point for creating ultracold molecules in the rovibrational ground state~\cite{Danzl2008}.

\section{Cs BEC production}

\begin{figure}
	\centering
		\includegraphics[width=\textwidth]{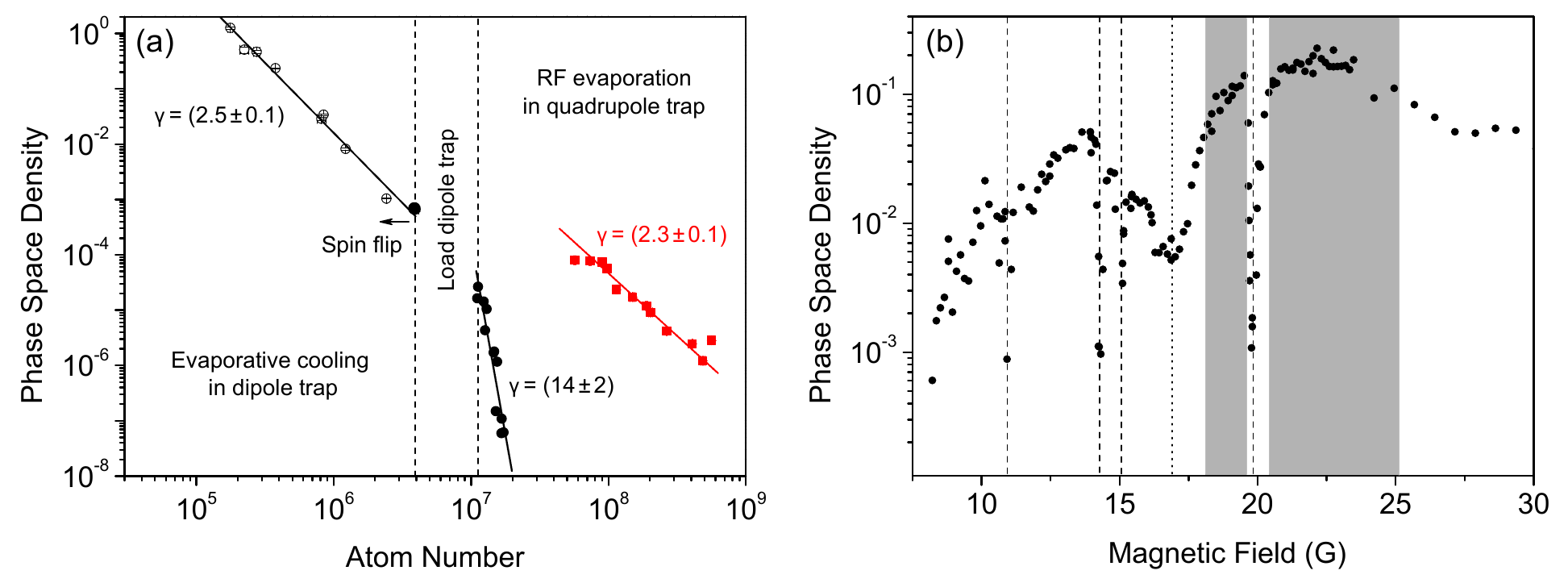}
	\caption{Evaporation efficiency. (a) Cs (black) and $^{87}$Rb (red) phase space density as a function of atom number throughout the experimental evaporation sequences in the magnetic and optical traps. Rb and Cs atoms are initially prepared in the $(1,-1)$ and $(3,-3)$ states respectively (solid symbols). Following the RF spin flip, the Cs atoms are transfered into the (3,+3) state (open symbols). Efficiency, $\gamma$, for each stage is shown, calculated according to equation~\ref{eqn}. (b) Phase space density after a fixed evaporation ramp as a function of magnetic field. The grey shaded regions indicate the field ranges in which Cs BECs can be created. Dashed lines show the experimentally determined positions of Feshbach resonances. The dotted line shows the position of the zero-crossing of the scattering length. }
	\label{fig:EvapEfficiency}
\end{figure}

We create Cs condensates using a sympathetic cooling scheme similar to that developed in our production of dual species Rb--Cs BECs \cite{Cho2011,McCarron2011a}, where $^{87}$Rb atoms are used as the refrigerant species. Below we present an outline of our method, focussing on the changes to the routine implemented when producing single species Cs condensates. We begin by loading a magnetic quadrupole trap with a Rb--Cs mixture from a dual species magneto-optical trap (MOT). Prior to loading, the Rb and Cs are optically pumped into the $(1,-1)$ and $(3,-3)$ states, respectively. Forced RF evaporation is then employed, with the choice of spin states effectively making the quadrupole trap deeper for Cs by a factor of three allowing the preferential removal of Rb atoms during the evaporation. The large interspecies scattering length~\cite{Jenkin2011} means the Cs atoms then rapidly equilibrate to the same temperature as the Rb atoms. This results in highly efficient evaporation of Cs as shown in figure~\ref{fig:EvapEfficiency}~(a). The evaporation sequence in the magnetic trap is initially optimised for $^{87}$Rb alone. The same sequence is used for the Rb--Cs mixture as only $\sim2\times10^7$ Cs atoms (compared to $\sim5\times10^8$ Rb atoms) are loaded into the trap thus the additional heat load has little impact on the evaporation trajectory. Note, when creating dual species condensates only $\sim3\times10^6$ Cs atoms are loaded initially~\cite{McCarron2011a}. We characterise the trajectory by the evaporation efficiency,
\begin{equation}
\gamma = \frac{\log ({\rm{PSD}}_2 / {\rm{PSD}}_1)} {\log(N_1/N_2)},
\label{eqn}
\end{equation}
where $\rm{PSD}_{1,2}$ are the phase space densities of the gas before and after the evaporation stage and $N_{1,2}$ are the initial and final atom numbers. This is shown for all stages of the evaporation in figure~\ref{fig:EvapEfficiency}~(a). In the magnetic trap, the sympathetic cooling of the Cs has an efficiency of 14(2) as relatively few atoms are lost whilst the temperature decreases. The evaporation in the magnetic trap results in a sample of $5.7(1) \times 10^7$ Rb atoms at a PSD of $7.9(9) \times 10^{-5}$ and $1.13(6) \times 10^7$ Cs atoms at a PSD of $2.6(3) \times 10^{-5}$. At this point Majorana spin flips limit the efficiency of any further cooling in this trap and so we transfer the atoms into a crossed optical dipole trap.

The crossed dipole trap~\cite{Jenkin2011} is formed from two independent beams derived from a single-frequency, $\lambda=1550$\,nm IPG Photonics ELR-30LP-SF fibre laser, each focussed to a waist of 68(1)\,$\mu$m. The beams intersect at an angle of 22\,$^\circ$ and are aligned $\sim~100\,\mu$m below the magnetic trap centre. With 6~W in each beam this results in a trap depth of $\sim$~100\,$\mu$K. Although the optical trapping beams are switched on throughout the RF evaporation the final transfer into the dipole trap only occurs when the vertical magnetic field gradient is decreased from 187\,G/cm to 28\,G/cm. The atoms are no longer completely levitated at this gradient and the optical trap is loaded through elastic collisions \cite{Jenkin2011}. During the transfer, the majority of the  $^{87}$Rb atoms is removed using the RF knife. Any residual $^{87}$Rb atoms are removed from the trap using a pulse of resonant light.

Typically $3.9(2) \times 10^6$ Cs atoms are loaded into the dipole trap at a temperature of 20.1(1)\,$\mu$K and a PSD of $6.5(5) \times 10^{-4}$. To suppress two-body losses \cite{Stamper-Kurn1998}, the atoms are transferred into their absolute ground state via RF rapid adiabatic passage \cite{Bergmann1998} whilst at the same time a magnetic bias field is added in such a way to produce a magnetic potential where the magnitude of the field increases in the upwards direction, thereby levitating the final high-field-seeking states of the atoms against gravity. This state transfer is essential for efficient cooling of Cs \cite{Soding1998,Thomas2003}. Following the spin flip, we are left with an atomic sample of $2.4(3) \times 10^6$ atoms in the (3,+3) state at a temperature of 15.04(4)\,$\mu$K and a PSD of $1.03(2) \times 10^{-3}$.

Despite the atoms being in the absolute ground state, three-body recombination can still hinder the progress of evaporation. Feshbach resonances are commonly employed to improve evaporation conditions \cite{Marchant2012}, effectively controlling the ratio of good (elastic) to bad (inelastic and reactive) collisions. The broad resonance in the (3,+3) state at 17\,G, shown in figure~\ref{fig:EvapEfficiency}~(b), is ideal for such an application, giving precise tuning of the scattering length.
A minimum in the three-body loss rate exists at $B \approx 21$\,G, offset from the minimum in the elastic scattering rate. This is the result of an Efimov effect \cite{Kraemer2006} and allows three-body losses to be minimised whilst still ensuring efficient rethermalisation of the atoms.

To probe the scattering properties of the atomic gas, and hence determine the optimum field for evaporative cooling, a fixed evaporation sequence was carried out at a range of magnetic fields. Figure~\ref{fig:EvapEfficiency}~(b) shows the phase space density of the resultant cloud over the region of 5 to 30\,G. We are able to locate a number of resonances over this field range, each characterised by a drop in atom number and an increase in temperature which results in a sharp decrease in phase space density, shown by dashed lines on figure~\ref{fig:EvapEfficiency}~(b). In addition a broad drop in the PSD at 16.9(1)\,G results from the drop in the elastic collision rate near the zero crossing in the scattering length. This is indicated by a dotted line on figure~\ref{fig:EvapEfficiency}~(b).

We find it is possible to condense Cs in two bias field regions (marked in grey on figure~\ref{fig:EvapEfficiency}~(b)), the first from 18.1 to 19.6\,G and the second from 20.2 to 25\,G. This corresponds to scattering length ranges from 60 to 160\,$a_0$ and 170 to 400\,$a_0$ respectively. These two regions are separated by a Feshbach resonance at 19.89\,G, later used for magnetoassociation. In the optimum case, evaporation to BEC is carried out at 22.91(2)\,G where the scattering length is $\sim310\,a_0$. Here, three linear ramps of the dipole trap power are executed, reducing the trap depth to 144~nK. At this point evaporation by merely lowering the optical trap depth stagnates and we instead proceed by tilting the trap~\cite{Hung2008} via an increase in field gradient to 33.6\,G/cm. This method produces a condensate of $1 \times 10^5$ Cs atoms. In the lower field region the evaporation sequence is slightly modified. Following initial ramps to decrease the beam powers, the magnetic gradient is rapidly changed to 42.3\,G/cm. This equalises the trap depths in both the vertical direction and along the beams. From here the gradient is ramped slowly, over 17.5\,s, to 62\,G/cm, thus tilting the trapping potential. This reduces the trap depth with little impact on the trap frequencies \cite{Hung2008,Jenkin2011}, maintaining the elastic collision rate for lower scattering lengths and resulting in a BEC with up to $3 \times 10^4$ atoms.


\section{Association of Cs$_2$ molecules}

\begin{figure}
	\centering
\includegraphics[width=0.5\textwidth]{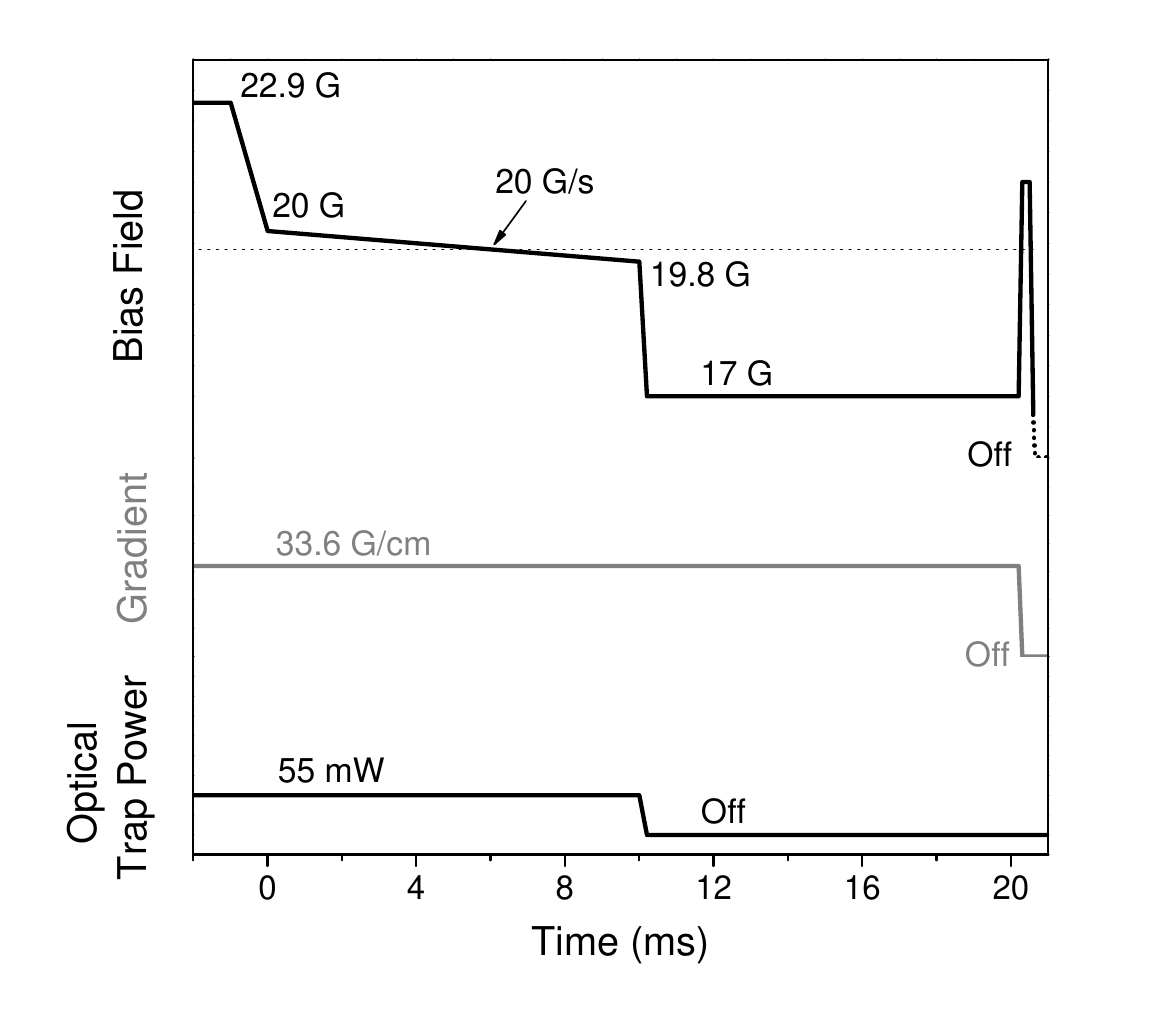}
	\caption[Association Sequence]{The experimental sequence for the magnetoassociation, Stern-Gerlach separation, and dissociation of Feshbach molecules after the production of an ultracold sample of Cs at 22.9\,G. The horizontal dotted line indicates the position of the Feshbach resonance at 19.89\,G used for the association. Shown are the bias field, magnetic gradient and dipole trap power. }
	\label{fig:CsAssociationSequence}
\end{figure}

\begin{figure}
	\centering
\includegraphics[width=0.75\textwidth]{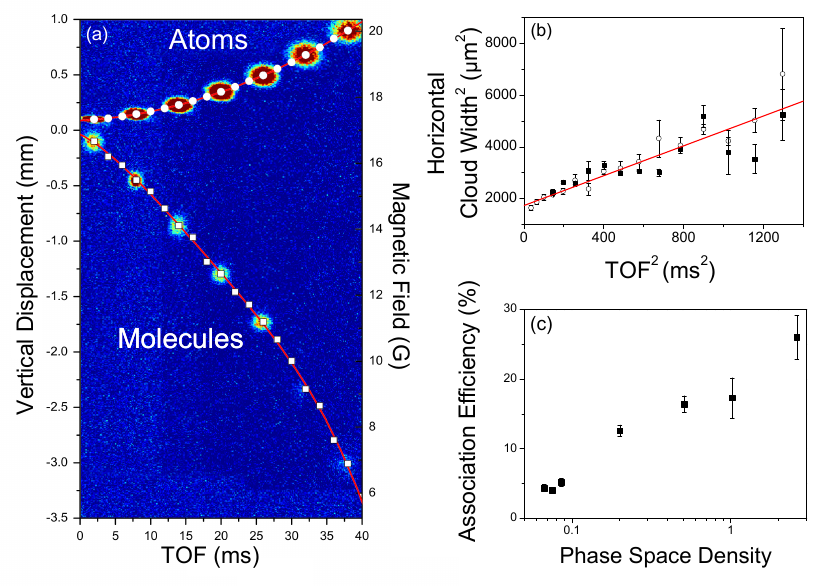}
	\caption{Association and Stern-Gerlach separation of the molecular cloud. (a) Absorption images taken with varying time of flight showing Stern-Gerlach separation of the atomic and molecular clouds. Overlayed are the measured mean positions of both the atomic and molecular cloud, along with fit curves to serve as a guide to the eye. The right axis shows the magnetic field experienced by the falling molecular cloud. Error bars on position are shown but are negligible at this scale. (b) The fitted widths of the atomic (open symbols) and molecular (filled symbols) clouds in the horizontal axis. The temperature of each cloud is determined by fitting curves generated by the numerical simulation of the molecular motion described in section~5 to the data, yielding 60(3)~nK for both the atoms and molecules. (c) The association efficiency of the process is observed to increase with phase space density, with a maximum efficiency of 26(3)\,\%  obtained.}
	\label{fig:SternGerlach}
\end{figure}

Following the successful creation of a high phase space density sample, magnetoassociation using the $4g(4)$ resonance at 19.89\,G is carried out to produce Cs$_2$ molecules. The experimental sequence after evaporation is shown in figure~\ref{fig:CsAssociationSequence}. Firstly, the magnetic field is decreased from 22.91\,G to a field just above the Feshbach resonance. The molecules are then associated by ramping over a 200\,mG window, centred on the resonance, in a period of 10\,ms. The resonance width determined from loss measurements is 0.10(2)\,G.  The subsequent ramp to 17\,G binds the molecules by 2\,MHz$\times h$. The bound molecules have a magnetic moment of $-0.9$\,$\mu_{\rm{B}}$, which corresponds to a levitation gradient of 52\,G/cm, compared to 31\,G/cm for the single Cs atoms. This difference means Stern-Gerlach separation can be used to isolate the two components for imaging. As we cannot image the molecules directly, they must first be dissociated. This is done by sweeping the magnetic field across the Feshbach resonance in the opposite direction to the association ramp. The dissociated atoms are then imaged using standard resonant absorption imaging following the turn off of the magnetic field. Figure~\ref{fig:SternGerlach} illustrates how the separation can be used to experimentally determine the temperature and magnetic moment of the atomic and molecular clouds. However, in this case the molecules fall through a sufficient range of magnetic field that several avoided crossings in the bound state spectrum (figure ~\ref{fig:CsScatBound30G} (b)) are traversed, altering the magnetic moment of the molecules during their motion. Consequently, the magnitude of the upwards force is not constant with position and the cloud trajectory cannot be used to reliably measure the magnetic moment of a given molecular state. This quantity is instead measured by finding the magnetic field gradient which exactly levitates the molecular cloud.

The absolute molecule number produced and association efficiency depend on the PSD of the atomic sample before association \cite{Hodby2005}, as shown in figure~\ref{fig:SternGerlach}(c). Here, the association efficiency is defined as $2N_{{\scriptsize \textrm{mol}}}/N_{{\scriptsize \textrm{Cs}}}$ where $N_{{\scriptsize \textrm{mol}}}$ is the number of molecules produced and $N_{{\scriptsize \textrm{Cs}}}$ is the initial number of Cs atoms. We are able to produce a maximum of $2.1(1) \times 10^4$ molecules, coupled out of a sample of $3.28(2) \times 10^5$ Cs atoms with an initial PSD of 0.20(1). We observe our highest association efficiency, 26(2)\,\%, when the molecules are coupled out of a BEC.



\section{Controlling the molecular state}

\begin{figure}
	\centering
\includegraphics[width=0.9\textwidth]{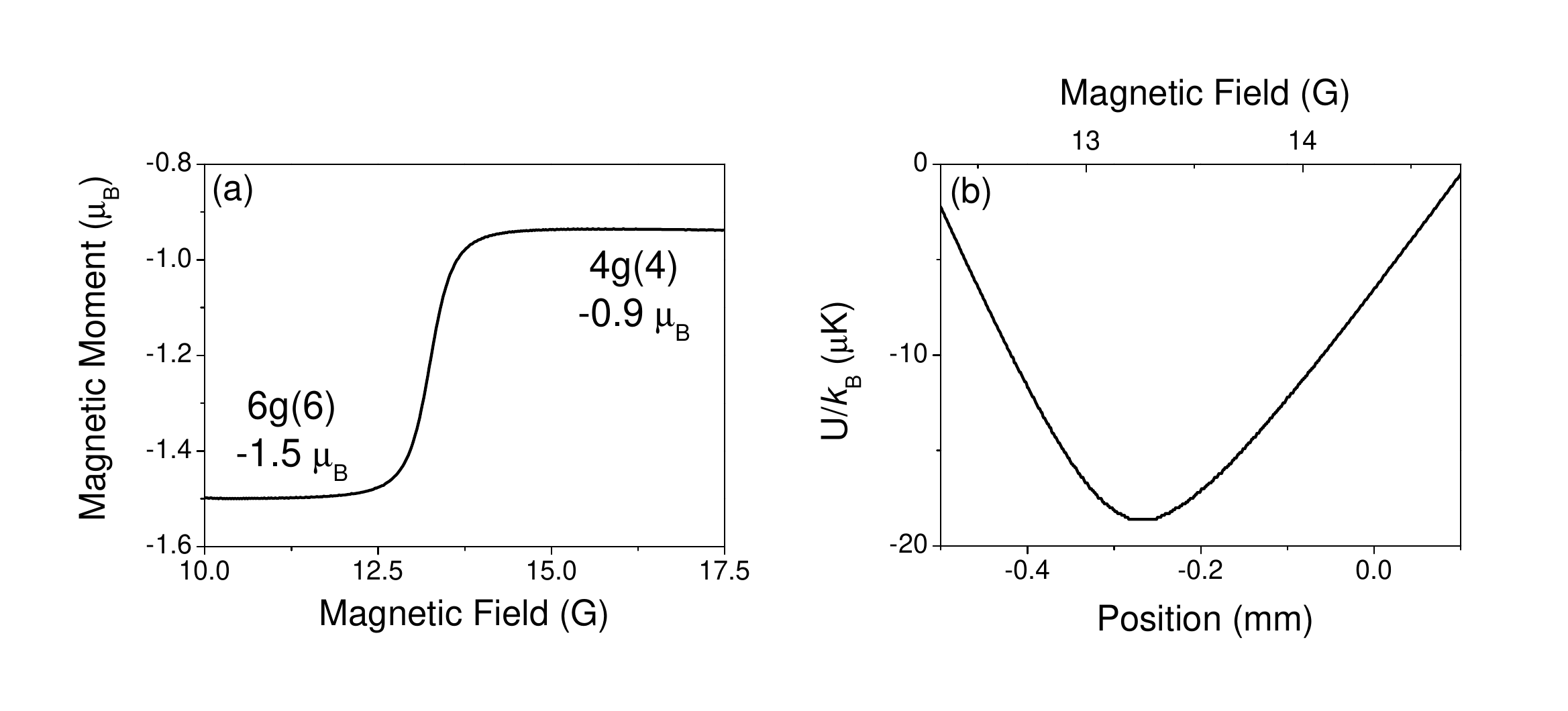}
	\caption[Magnetic moment and potential.]{Behaviour of the magnetic moment over the avoided crossing at $\sim13.3$\,G. (a) Magnetic moment versus bias field of a Cs$_2$ dimer adiabatically following the avoided crossing between the $4g(4)$ and $6g(6)$-state. (b) The potential for the Cs$_2$ molecules for a magnetic field gradient of 40\,G/cm in the vertical direction resulting from their change in magnetic moment with position. The effect of gravity is also included.}
	\label{fig:MagnMom}
\end{figure}

In the bound-state spectrum, shown in figure~\ref{fig:CsScatBound30G}, one can see that avoided crossings between different molecular states exist. These can be used to change the internal state of the molecules by changing the magnetic bias field \cite{Mark2007}, thereby controlling their magnetic moment. Below we focus on the first avoided crossing of the $4g(4)$-state which occurs with the $6g(6)$-state at $\sim13.3$\,G, shown in figure~\ref{fig:MagnMom}~(a). For this avoided crossing the coupling strength between the two states is $V \cong h \times 150$\,kHz \cite{Chin2005}.

Using the bound-state spectrum it is possible to calculate the magnetic moment of the molecules. In this bound-state picture the molecular binding energy is given with respect to the energy of two unbound atoms which changes with magnetic field according to the Breit-Rabi formula~\cite{Breit1931}. The gradient of the absolute bound-state energy gives the magnetic moment of the molecules. The magnetic moment does not change instantaneously between states, but follows the curvature of the avoided crossing between them. Figure~\ref{fig:MagnMom}~(a) shows how the magnetic moment of a molecular sample adiabatically follows the avoided crossing between the $4g(4)$ and $6g(6)$-states at 13.3\,G.

To implement a change of state it is necessary to precisely control Landau-Zener tunelling at an avoided crossing. The rate of the magnetic field ramps used determines whether or not the crossing is adiabatically followed. In the Landau-Zener model avoided crossings are characterised by two parameters: the coupling strength $V$, which corresponds to half the energy splitting between the two states at the crossing point, and $\Delta \mu$, the difference in magnetic moment $\mu = -dE/dB$ of the two states. The critical ramp speed, $\dot{r}_{\rm{c}}$, is defined as \cite{Mark2007}

\begin{equation}
\dot{r}_{\rm{c}}=\frac{2\pi V^2}{\hbar |\Delta\mu|}.
\label{eq:AdiabSweep}
\end{equation}
For fast magnetic field ramps where the rate of change in magnetic field, $\dot{B} \gg \dot{r}_{\rm{c}}$ the crossing is diabatic and the population remains in the same state. However, for slow ramps, $\dot{B} \ll \dot{r}_{\rm{c}}$, the population is transferred adiabatically to the other state.

The critical ramp speed for the states of interest is $\dot{r}_{\rm{c}} \sim 1100$\,G/ms. The $6g(6)$-state has a magnetic moment of $-1.5$\,$\mu_{\rm{B}}$ and hence a levitation \mbox{gradient of 31\,G/cm}. The difference in magnetic moment of the two states involved in this avoided crossing makes it possible to apply a magnetic field gradient that over-levitates one state and under-levitates the other. Combined with the fact that the molecules experience a different bias field as a function of vertical position due to the magnetic field gradient (see figure~\ref{fig:SternGerlach} (a)), this gives us the ability to manipulate the internal state of the molecular cloud as it falls. Figure~\ref{fig:MagnMom}(a) shows the variation of the magnetic moment with magnetic field as the avoided crossing between the two states is traversed adiabatically. We expect that molecules released from the optical trap will initially fall before beginning to oscillate in the vertical direction as the molecules are transferred between the two states~\cite{Durr2004a}.

\begin{figure}
	\centering
\includegraphics[width=1\textwidth]{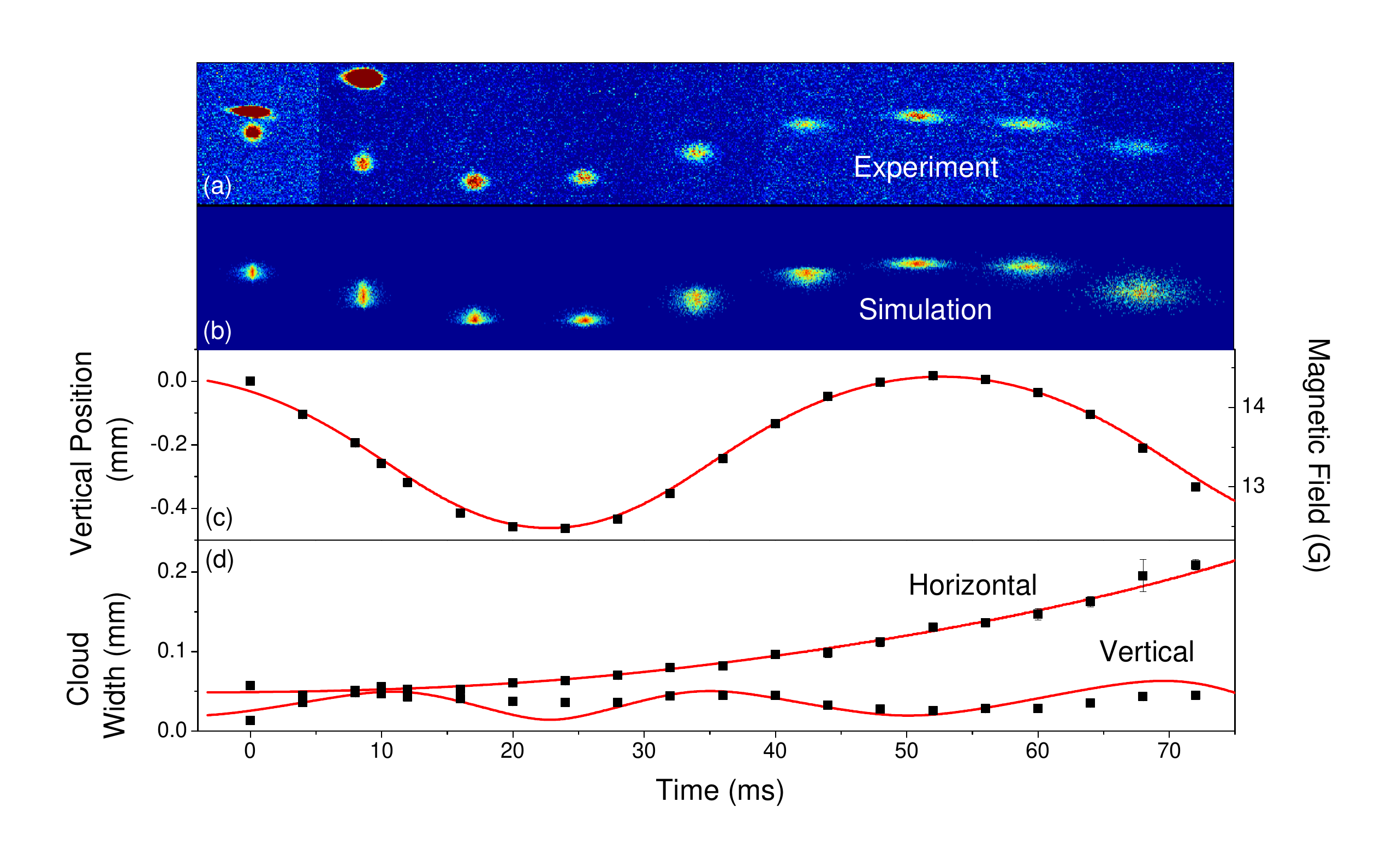}
	\caption[Fit to 'bouncing' molecules.]{Position and width of the molecular cloud versus time. Zero time corresponds to the end of the magnetic field ramp up to 40~G/cm, such that the molecules are already separated from the atomic cloud. (a) Experimental absorption images taken at 8~ms intervals and (b) results of the theoretical simulation. The motion of the molecules is simulated in a quadrupole field with a magnetic field gradient of 40\,G/cm using the potential shown in figure~\ref{fig:MagnMom}. (c) Vertical position of the cloud as a function of time. The solid line is the output of the simulation, showing good agreement with the data. (d) Cloud widths in the horizontal and vertical directions during the `bouncing' and the corresponding theoretical values (solid lines). }
	\label{fig:PosFit}
\end{figure}

To demonstrate this experimentally we begin by producing a sample of Feshbach molecules. Following the Feshbach association at 19.89\,G the bias field is rapidly changed to 14.9\,G at the position of the optical dipole trap. At the same time the magnetic field gradient is ramped to 40\,G/cm. Once the optical trap is switched off the molecules begin to fall through the field gradient, in the direction of decreasing field. When the magnitude of the net field drops just below 13.3\,G the molecules are transferred into the $6g(6)$-state and are pushed upward as a result of the over-levitation of this state, appearing to `bounce'. As the molecules travel upwards and move to a region of higher field, they are transferred back into the $4g(4)$-state. As the gradient field is insufficient to support the molecules in this state they begin to fall under gravity once more. Figure~\ref{fig:PosFit}~(a) shows absorption images of the `bouncing' molecules, clearly demonstrating the vertical oscillation of the cloud. Note that the magnetic field ramp to 14.9~G is important to ensure that the molecular cloud begins its trajectory sufficiently close to the avoided crossing. If the cloud is falling too quickly as it enters the $6g(6)$-state, the upwards force provided is not sufficient to change the direction of motion before the neighboring crossing to the $4g(3)$-state is encountered.

We use Monte-Carlo methods to simulate the motion of a cloud of 5000 non-interacting molecules in the experiment, moving in the applied magnetic field. The magnetic quadrupole field alone generates a linear potential in all directions, ${\vec{B}=B'_xx\vec{\hat{x}} + B'_yy\vec{\hat{y}} +B'_zz\vec{\hat{z}}}$, where, in accordance with Maxwell's equations, the gradient $B'\equiv B'_x = B'_y = -B'_z/2$. The addition of the bias field, $B_0\vec{\hat{z}}$, along the $z$-axis results in a linear potential chosen to exactly cancel the effect of gravity (for certain states) whilst a weak harmonic potential is created along the $x$- and $y$-directions. In our model we take into account the exact coil configurations used in the experimental apparatus, calculating the magnetic field on a 3D matrix of points in space. We find the trajectory of each molecule in the cloud by numerically solving the classical system of equations of motion in cylindrical coordinates,

\begin{equation}
\frac{{d}^{2} z}{{d}t^{2}} = -\frac{\vec{\mu} (z,r)}{m} \cdot \frac{\partial \vec{B} (z,r)}{\partial z} - g, \\
\frac{{d}^{2} r}{{d}t^{2}} = -\frac{\vec{\mu} (z,r)}{m} \cdot \frac{\partial \vec{B} (z,r)}{\partial r}, \\
\label{eq:motion}
\end{equation}
where $m$ is the mass of the molecules, $g$ the gravitational acceleration and $z$ and $r$ are the vertical and radial displacements respectively. The magnetic field we apply is cylindrically symmetric about the $z$ axis, the azimuthal angle $\phi$ is therefore obtained trivially by conservation of angular momentum. As the maximum rate of change of the magnetic field during the motion is 0.134\,G/ms, well below the critical ramp speed of $\sim 1100$\,G/ms (see equation~\ref{eq:AdiabSweep}), we assume the molecules are all transferred adiabatically between the $4g(4)$ and $6g(6)$-states, so that the magnetic moment follows that in figure~\ref{fig:MagnMom}~(a). As a consequence, the total potential of the molecules $U = \vec{\mu} \cdot \vec{B} + gz$ shows a pronounced minimum in the vertical direction (figure~\ref{fig:MagnMom}~(b)) despite the fact that there is only a linear gradient in the magnetic field. It is motion in this potential that leads to the `bouncing' shown in figure~\ref{fig:PosFit}~(a).

We define the initial spatial distribution of the molecules to mirror the distribution of the atoms in the cloud from which they are associated. The cloud is therefore specified by a 3-dimensional Gaussian distribution with a narrower width in the vertical plane, such that the cloud is ellipsoidal in shape, reflecting the aspect ratio of the optical trap. The initial velocity distribution of the molecules is thermal, and hence given by the Maxwell-Boltzmann distribution. We use the experimentally measured cloud temperature of 60(3)~nK in the simulation, calculated from figure~\ref{fig:SternGerlach}(b). In addition, the molecules are not all associated simultaneously, but over a range of times specified by the width of the atomic cloud and the width of the Feshbach resonance which dictates the speed of the association ramp. As a result, molecules at the top of the cloud begin to fall before those at the bottom, which we account for as an extra contribution to the initial spatial and velocity distributions.

The comparison of the results of the theoretical simulation to the experiment is shown in figure\,\ref{fig:PosFit}. Figures\,\ref{fig:PosFit}\,(a) and (b) show absorption images of the oscillating Cs$_2$ cloud and the simulated model. Figures\,\ref{fig:PosFit}\,(c) and (d) show the simulated cloud position and widths in the horizontal and vertical directions. We find excellent agreement between the theoretical model and the experimental results.



\section{Repeated output coupling of molecules}

\begin{figure}
	\centering
\includegraphics[width=0.9\textwidth]{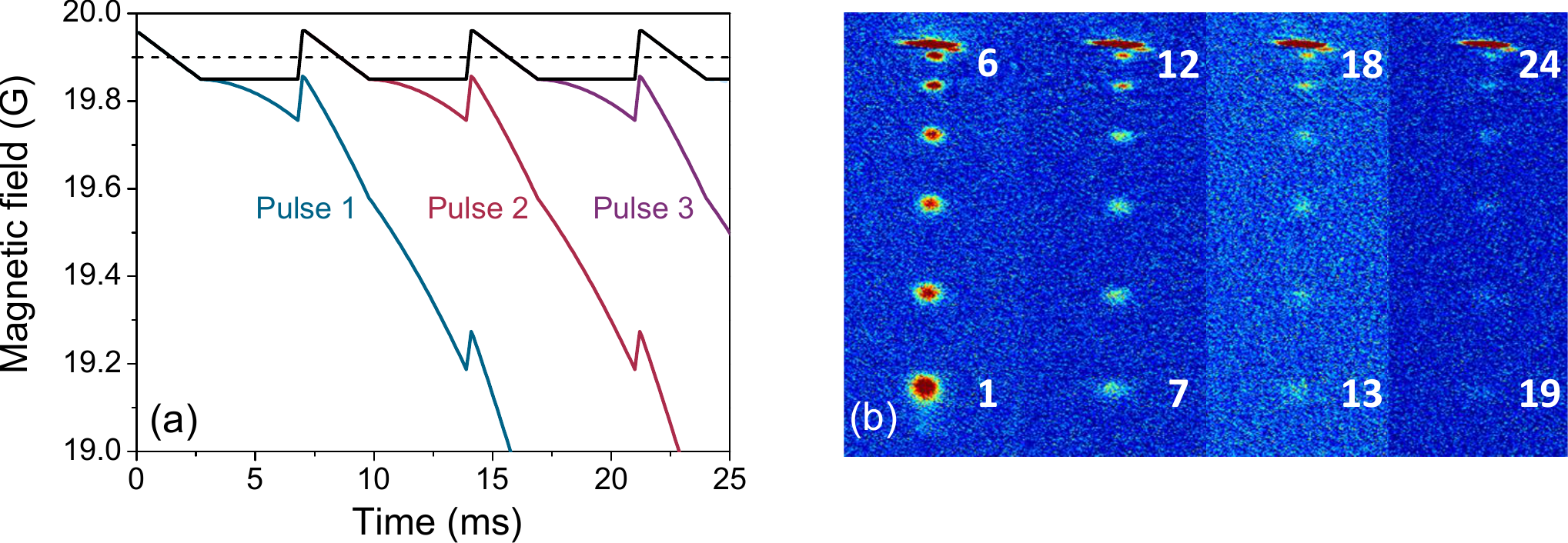}
	\caption[Magnetic field during Stern-Gerlach separation.]{Multiple output coupling of molecules (a) The magnetic field at the position of the trapped atoms (black) during the repeated output coupling sequence. Each pulse of output coupled molecules falls under gravity and therefore experiences a different magnetic field (coloured trajectories). The dashed line marks the position of the Feshbach resonance. Crucially, the magnetic field at the position of the atoms can be jumped back across the Feshbach resonance without dissociating the molecules, thus multiple samples of molecules can be produced from a single atomic cloud. (b) Absorption images of the molecular samples. The atomic cloud in the dipole trap can be seen on the top of each image with the falling molecular pulses below. Only 6 pulses can be imaged at any one time as the molecules fall out of the field of view of the camera. The size of each image is $4.35 \times 1.65$\,mm.}
	\label{fig:Multiploutput}
\end{figure}

It has been shown~\cite{Mewes1997} that atoms can be controllably and repeatable output coupled from an atomic source allowing the realisation of a pulsed `atom laser'. In the experimental setup presented here, the presence of the magnetic field gradient during the Stern-Gerlach separation enables the selective addressing of atoms and molecules based on their position. This makes it possible to perform repeated output coupling of \textit{molecules} from a single atomic cloud, echoing the work on atom lasers. We first carry out the standard magnetoassociation sequence similar to that illustrated in figure~\ref{fig:CsAssociationSequence}, but with the optical dipole trap left on. The magnetic field gradient levitates the atoms but is insufficient to support the molecules. The molecules therefore escape the shallow dipole trap, despite the fact that the laser beams are left on throughout. The output coupled molecules then fall under gravity towards the region of lower field, as shown in figure\,\ref{fig:SternGerlach}\,(a). Once the molecules have dropped sufficiently far, it is possible to increase the bias field back to a value above the Feshbach resonance at the position of the atoms in the trap without dissociating the molecules. Another molecular sample can then be produced from the atomic cloud by sweeping across the resonance again. This process can be repeated multiple times.

The magnetic field ramping sequence and the field the atoms and molecules are exposed to are shown in figure~\ref{fig:Multiploutput}~(a). The association ramp in this sequence is a magnetic field sweep from 19.96\,G to 19.85\,G in 2.7\,ms, around twice as fast as our usual ramp. The field is then held at 19.85\,G for 4.1\,ms until the molecular cloud has separated sufficiently from the atomic cloud. After this, the field at the position of the atoms is jumped back across the resonance to 19.96\,G in 0.2\,ms. After a hold time of 0.1\,ms, the field is again ramped down to 19.85\,G in 2.7\,ms to produce another sample of molecules. In this way, a pulse of Cs$_2$ molecules can be produced every 7.1\,ms. Molecules were detected for up to 24 of these association ramps as shown in figure~\ref{fig:Multiploutput}~(b). Only six molecular samples can be imaged at one time due to the limited field of view of the camera. The top cloud present in all the images is the Cs atoms confined in the dipole trap.

\begin{figure}
	\centering
				\includegraphics[width=\textwidth]{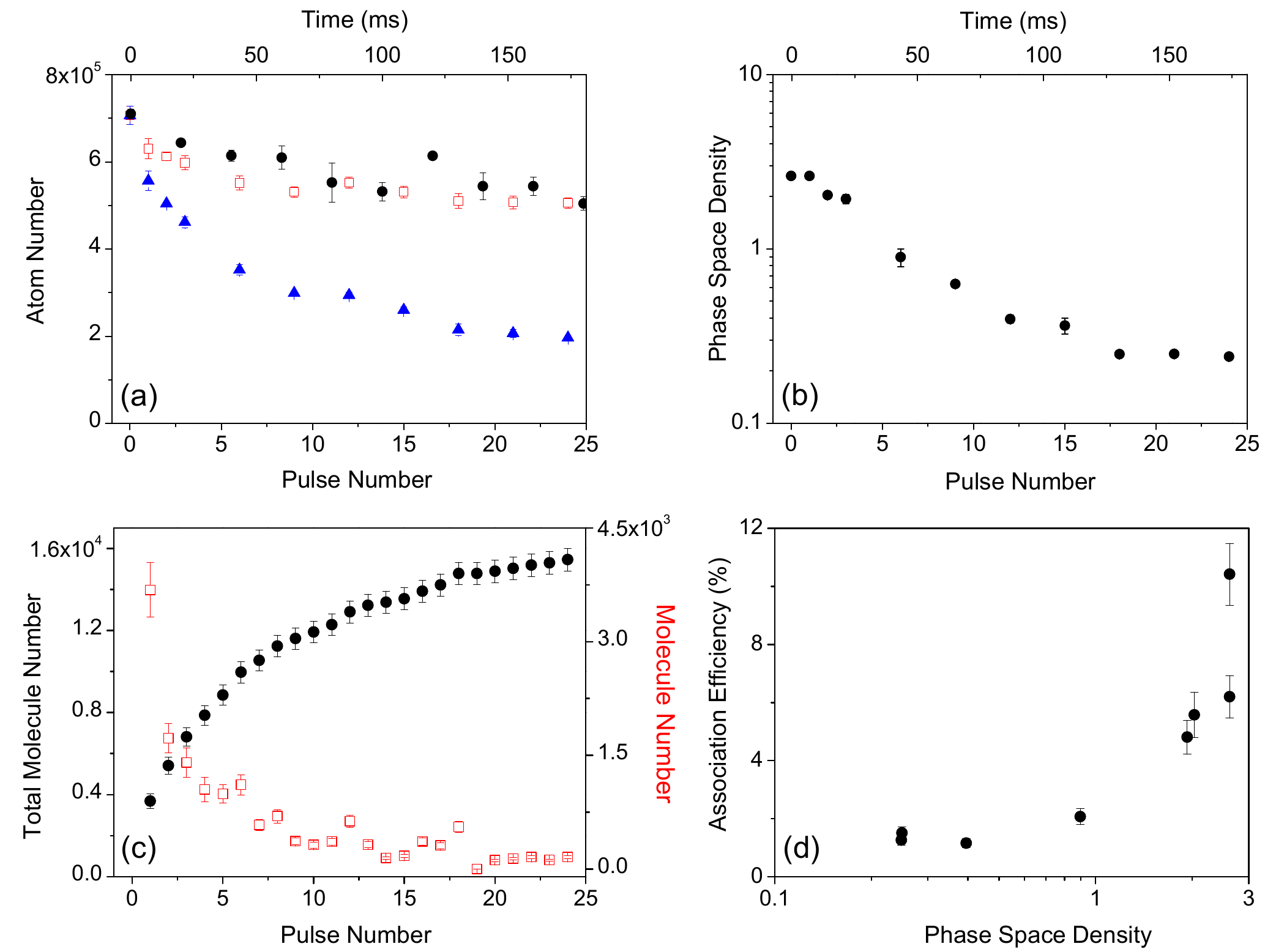}
	\caption{Repeated output coupling of molecules. (a) Blue triangles show the number of atoms remaining in the initial trapped cloud as molecules are coupled out. The atom number in the trap without molecule production is shown by the black circles. The red squares represent the result of summing the number of atoms left in the trap after molecule production and the number of atoms converted into molecules. As the summed atom number and the unperturbed number agree, we conclude that trap loss is solely due to the output coupling sequence.  (b) Evolution of the atomic phase space density as molecules are coupled out of the cloud. (c) Integrated number of molecules produced (black circles) and the number of molecules associated in each association ramp (red squares). (d) Association efficiency during the multiple output coupling sequence. The efficiency with respect to PSD decreases faster than that observed in the single pulse production (see figure~\ref{fig:SternGerlach}~(c)) indicating that the atomic gas does not fully rethermalise between output pulses.   }
	\label{fig:Mol_production}
\end{figure}

The molecule production in this sequence is less efficient than in the single pulse production sequence owing to the faster ramp time across the resonance needed to achieve a high repetition rate. In figures~\ref{fig:Mol_production}~(a) and (b) the effect of the output coupling of molecules on the atomic sample is shown. By measuring the number of atoms remaining in the trap and the number associated into molecules we are able to infer that the molecule production is the only source of atom loss from the trap on the timescale of the association. Each molecular pulse decreases the PSD of the atomic cloud (shown in figure~\ref{fig:Mol_production}~(b)) hence, in order to increase the total number of molecular pulses that can be coupled out of a single atomic sample, the atoms are cooled to BEC before the out coupling sequence to maximise the PSD. In the 24 pulse sequence the first two molecular samples are coupled out of a Cs BEC.

In figure~\ref{fig:Mol_production}~(c), the molecule number in each pulse and the cumulative total of molecules from a single atomic sample are shown. In the first association ramp, $3.7(4) \times 10^3$ molecules are produced. Following 24 successive ramps the number of molecules out coupled from the same atomic sample totals $1.55(6) \times 10^4$. As more molecules are coupled out of the atomic cloud, the phase space density drops and so, in turn, the association efficiency decreases. This dependence is illustrated in figure~\ref{fig:Mol_production}~(d). Note that the efficiency with respect to PSD decreases faster than that observed in the single pulse production (see figure~\ref{fig:SternGerlach}~(c)) indicating that the atomic gas does not fully rethermalise between output pulses. This raises the possibility to use the repeated output coupling of molecules to probe the time evolution of the PSD and hence rethermalisaion of the atomic cloud. By varying the time between two molecular pulses and observing the association efficiency of the second pulse, one could directly probe the rethermalisation of the cloud following the initial magnetoassociation.



\section{`Collision' of two molecular samples}

\begin{figure}
	\centering
\includegraphics[width=0.9\textwidth]{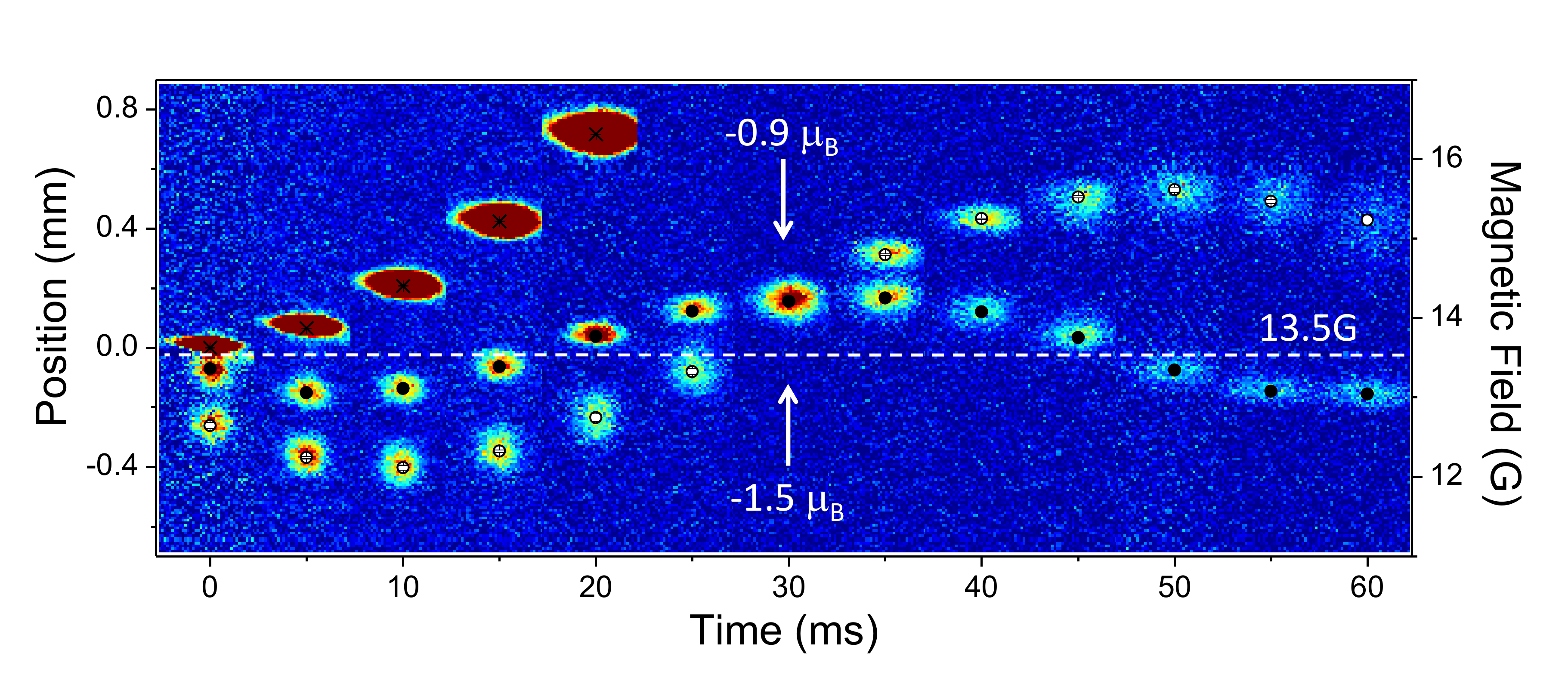}
	\caption[Cs$_2$ `collision'.]{Bouncing and `collision' of two molecular clouds. Two molecular clouds are coupled out of one atomic sample and `bounce' in a gradient field of 40\,G/cm. The outcoupling and bias fields are chosen such that the two molecular clouds `collide' after 30\,ms time of flight. }
	\label{fig:ColPic}
\end{figure}

By combining the techniques of multiple output coupling and molecule `bouncing' it is possible to engineer a `collision' between two molecular clouds.
To begin two molecular clouds are coupled out of the atomic sample. After the second output coupling process, the magnetic bias field is ramped down to transfer the molecules into the $6g(6)$-state. At the same time, the field gradient is increased to 40\,G/cm such that the molecules experience a force pushing them upwards. The trajectory of the two clouds in this net field is shown in figure~\ref{fig:ColPic}. The two clouds oscillate in the potential shown in figure~\ref{fig:MagnMom}~(b), each with different amplitudes, set by their position and velocity at the time that the field is switched. At 30\,ms the two clouds reach the same point in the vertical direction and overlap. From simulations of the positions and the widths of the molecular clouds similar to the ones presented in figure~\ref{fig:PosFit}, we conclude the motion of the two molecular clouds is unperturbed by this `collision'. We attribute this to the low molecular density of $\sim 10^8$\,cm$^{-3}$ in the present experiment. Nevertheless the results illustrate the exquisite control with which the molecular clouds can be manipulated and potentially demonstrate a route towards a low energy molecular collider, as realised with atomic samples~\cite{Thomas2004a, Buggle2004}, or the accumulation of multiple molecular pulses in a second trap. Moreover, since the two molecular clouds are coupled out of the same Cs BEC it could be possible to observe interference between them in future experiments, similar to the interferences observed between two BECs \cite{Andrews1997}.

\section{Conclusion}

In summary we have demonstrated the repeated output coupling of ultracold Feshbach molecules from a single Cs atomic cloud. Bose—-Einstein condensates of Cs containing up to $1\times10^5$ atoms are produced in a crossed optical dipole trap. We characterise the magnetoassociation of ultracold Feshbach molecules using the $4g(4)$ resonance in the (3,+3) state of Cs at 19.89~G. We create up to $2.1(1)\times10^4$ Feshbach molecules at a temperature of $\sim60(3)$~nK in a single association sequence. By exploiting the motion of the falling molecular clouds in the presence of a magnetic field gradient we are able to demonstrate the repeated output coupling of molecules from a single atomic sample. The presence of a strong avoided crossing in the bound state spectrum between the $4(g)4$-state used for magetoassociation and a $6g(6)$-state at 13.3~G provides a further mechanism to control the dynamics of the molecular clouds.

There are a number of directions for future experiments using the repeated output coupling sequence. Increasing the density of the molecular sample should allow us to probe ultracold collisions between Feshbach molecules as a function of collision energy. The use of a second dipole trap may allow the accumulation of multiple molecular pulses in a single trap. Combining dynamic magnetic field control with the rich near threshold bound state structure may allow pulses of molecules to be overlapped and subsequently trapped by turning on the second dipole trap. The addition of further evaporative cooling of the trapped atomic gas between the output coupling pulses should improve the overall conversion efficiency. Finally the efficiency of the molecule production may be used as a microscopic probe of rethermalisation in the atomic cloud following magnetoassociation.


\section*{Acknowledgments}

We thank J. M. Hutson and C. L. Blackley for many valuable discussion and for providing theoretical bound-state energies and magnetic moments for Cs$_2$. This work was supported by the UK EPSRC.


\section*{References}


\providecommand{\newblock}{}

\end{document}